\documentclass[10pt,english,conference]{IEEEtran}
\usepackage[T1]{fontenc}
\usepackage[latin9]{inputenc}
\usepackage{color}
\usepackage{babel}
\usepackage{verbatim}
\usepackage{float}
\usepackage{amsthm}
\usepackage{amsmath}
\usepackage{amssymb}
\usepackage{graphicx}
\usepackage{setspace}
\usepackage{esint}
\usepackage{epstopdf}

\makeatletter

\pdfpageheight\paperheight
\pdfpagewidth\paperwidth

\providecommand{\tabularnewline}{\\}

\theoremstyle{plain}
\newtheorem{thm}{\protect\theoremname}
\theoremstyle{plain}
\newtheorem{lem}[thm]{\protect\lemmaname}

\usepackage{subfigure}
\usepackage{cite}
\usepackage{citesort}
\usepackage{balance}

\makeatother

\providecommand{\lemmaname}{Lemma}
\providecommand{\theoremname}{Theorem}

\begin{document}

\title{Approximation of Uplink Inter-Cell Interference in FDMA Small Cell
Networks
}

\begin{singlespace}

\author{\noindent {\normalsize{}Ming Ding, }\emph{\normalsize{}National ICT
Australia (NICTA), Australia}{\normalsize{} \{$\mathtt{Ming.Ding@nicta.com.au}$\}}\\
{\normalsize{}David L$\acute{\textrm{o}}$pez P$\acute{\textrm{e}}$rez,
}\emph{\normalsize{}Bell Labs Alcatel-Lucent, Ireland}{\normalsize{}
\{$\mathtt{dr.david.lopez@ieee.org}$\}}\\
{\normalsize{}Guoqiang Mao, }\emph{\normalsize{}The University of
Technology Sydney and NICTA, Australia}{\normalsize{} \{$\mathtt{Guoqiang.Mao@uts.edu.au}$\}}\\
{\normalsize{}Zihuai Lin, }\emph{\normalsize{}The University of Sydney,
Australia}{\normalsize{} \{$\mathtt{zihuai.lin@sydney.edu.au}$\}}}
\end{singlespace}

\maketitle
{}
\begin{abstract}
In this paper, for the first time, we analytically prove that the
uplink (UL) inter-cell interference in frequency division multiple
access (FDMA) small cell networks (SCNs) can be well approximated
by a lognormal distribution under a certain condition. The lognormal
approximation is vital because it allows tractable network performance
analysis with closed-form expressions. The derived condition, under
which the lognormal approximation applies, does not pose particular
requirements on the shapes/sizes of user equipment (UE) distribution areas as in previous
works. Instead, our results show that if a path loss related random
variable (RV) associated with the UE distribution area, has a low
ratio of the 3rd absolute moment to the variance, the lognormal approximation
will hold. Analytical and simulation results show that the derived
condition can be readily satisfied in future dense/ultra-dense SCNs,
indicating that our conclusions are very useful for network performance
analysis of the 5th generation (5G) systems with more general cell
deployment beyond the widely used Poisson deployment.
\footnote{1536-1276 © 2015 IEEE. Personal use is permitted, but republication/redistribution requires IEEE permission. Please find the final version in IEEE from the link: http://ieeexplore.ieee.org/document/7417160/. Digital Object Identifier: 10.1109/GLOCOM.2015.7417160}
\end{abstract}

\section{Introduction}

Small cell networks (SCNs) have been identified as one of the key
enabling technologies in the 5th generation (5G) networks~\cite{Tutor_smallcell}.
In order to gain a deep theoretical understanding of the implications
that SCNs bring about, %
new and more powerful network performance analysis techniques are
being developed. In this context, new performance analysis tools can
be broadly classified into two large groups, i.e., macro-scopic analysis
and micro-scopic analysis~{[}2-8{]}.

The macro-scopic analysis assumes that both user equipments (UEs)
and small cell base stations (BSs) are randomly placed in the network,
often following homogeneous Poisson distributions~{[}2,3{]}.%
{} The micro-scopic analysis is usually conducted assuming that UEs are
randomly dropped and the BSs are deterministically deployed, i.e.,
the BS positions in the considered cellular network are known. %
Generally speaking, the macro-scopic analysis investigates network
performance on a high level~{[}2,3{]}, while the micro-scopic analysis
gives more detailed results for specific networks~{[}4-8{]}. Note
that both analyses are related to each other. The average performance
of %
micro-scopic analyses conducted over a large number of realizations
of BS deployments should be equal to that of the macro-scopic analysis,
provided that the examined realizations of the deterministic BS deployments
follow the BS distribution assumed in the corresponding macro-scopic analysis.

In this paper, we focus on the micro-scopic analysis, and in particular,
we consider an uplink (UL) frequency division multiple access (FDMA)
SCN. %
Note that the analysis of a UL FDMA system is more involved than its
downlink counterpart due to the power control mechanisms used at UEs.
For the UL micro-scopic analysis, existing works either use
\begin{itemize}
\item Approach~1, which provides closed-form but complicated analytical
results for a network with\emph{ a small number of interfering cells,
each cell with a regularly-shaped UE distribution area}, e.g., a disk
or a hexagon~\cite{UL_interf_2cells_ICC};
\item Approach~2, which makes a empirical conjecture on the UL interference
distribution and on that basis derives analytical results for a network
with \emph{multiple interfering cells placed on a regularly-shaped
lattice}, e.g., a hexagonal lattice~\cite{UL_interf_LN_conjecture_CDMA},~\cite{UL_interf_LN_conjecture_WCL};
\item Approach~3, which conducts system-level simulations to directly obtain
empirical results for a complex network with \emph{practical deployment
of multiple cells placed on irregular locations}~\cite{Tutor_smallcell},~\cite{UL_interf_sim1},~\cite{UL_interf_sim2}.
\end{itemize}

Obviously, Approach~1 and Approach~3 lack generality and analytical
rigorousness, respectively. Regarding Approach~2, it has been a number
of years since an important conjecture was proposed and extensively
used in performance analysis, which stated that the UL inter-cell
interference with disk-shaped UE distribution areas could be well
approximated by a lognormal distribution in code division multiple access (CDMA) SCNs~\cite{UL_interf_LN_conjecture_CDMA}
and in FDMA SCNs~\cite{UL_interf_LN_conjecture_WCL}. This conjecture
is vital since it
allows tractable network performance analysis with closed-form expressions.
However, there are two intriguing questions regarding this conjecture:
\emph{(i) Will the approximation still hold for non-disk-shaped UE
distribution areas? (ii) Will it depend on the sizes of the UE distribution
areas?} In this paper, we aim to answer those two questions, and thus
making significant contributions to constructing a formal tool for
the UL micro-scopic analysis of network performance.

Compared with the previous works~{[}5,6{]} of the micro-scopic network
performance analysis based on empirical studies, the contributions
of this paper are as follows:
\begin{enumerate}
\item Our work, for the first time, analytically proves the conjecture in~\cite{UL_interf_LN_conjecture_WCL},
i.e., the UL inter-cell interference in FDMA SCNs can be well approximated
by a lognormal distribution under a certain condition.
\item The derived condition, under which the lognormal approximation applies,
does not rely on particular shapes/sizes of UE distribution areas
and can be readily satisfied in future dense/ultra-dense SCNs, indicating
that conclusions derived using this framework are very general and
useful for network performance analysis.
\item Based on our work, we propose a new approach to fill an important
theoretical gap in the existing micro-scopic analysis, which either
assumes very simple BS deployments or relies on empirical results.
Specifically, we directly investigate a complex network with \emph{practical
deployment of multiple cells placed on irregular locations}. In order
to do that we provide a theoretical framework based on the lognormal approximation
of the UL interference distribution, supported by rigorous theoretical
analysis, as well as the conditions under which the approximation
should be valid.
\end{enumerate}

\section{Network Scenario and System Model\label{sec:Network-Model}}

In this paper, we consider UL transmissions and assume that in one
frequency resource block (RB) at a time slot, only \emph{one} UE is
scheduled by each small cell BS to perform an UL transmission, which
is a reasonable assumption in line with the 4th generation (4G) networks,
i.e., the UL single-carrier FDMA (SC-FDMA) system and the UL orthogonal
FDMA (OFDMA) system in the 3rd Generation Partnership Project (3GPP)
Long Term Evolution (LTE) networks~\cite{TS36.213} and in the Worldwide
Interoperability for Microwave Access (WiMAX) networks~\cite{WiMAX_AI}, respectively.

Regarding the network scenario, we consider a SCN with multiple small
cells operating on the same carrier frequency as shown in Fig.~\ref{fig:sys_model}.
In Fig.~\ref{fig:sys_model}, a total of $B$ small cells exist in
the network, including one small cell of interest denoted by $C_{1}$
and $B-1$ interfering small cells denoted by $C_{b},b\in\left\{ 2,\ldots,B\right\} $.
We consider a particular frequency RB, and denote by $K_{b}$ the
active UE associated with small cell $C_{b}$ using that frequency RB.
Moreover, we denote by $R_{b}$\emph{ the UE distribution area} of
small cell $C_{b}$, in which its associated UEs are distributed.%

\noindent \begin{center}
\begin{figure}[H]
\vspace{-0.5cm}

\noindent \begin{centering}
\includegraphics[width=6.5cm]{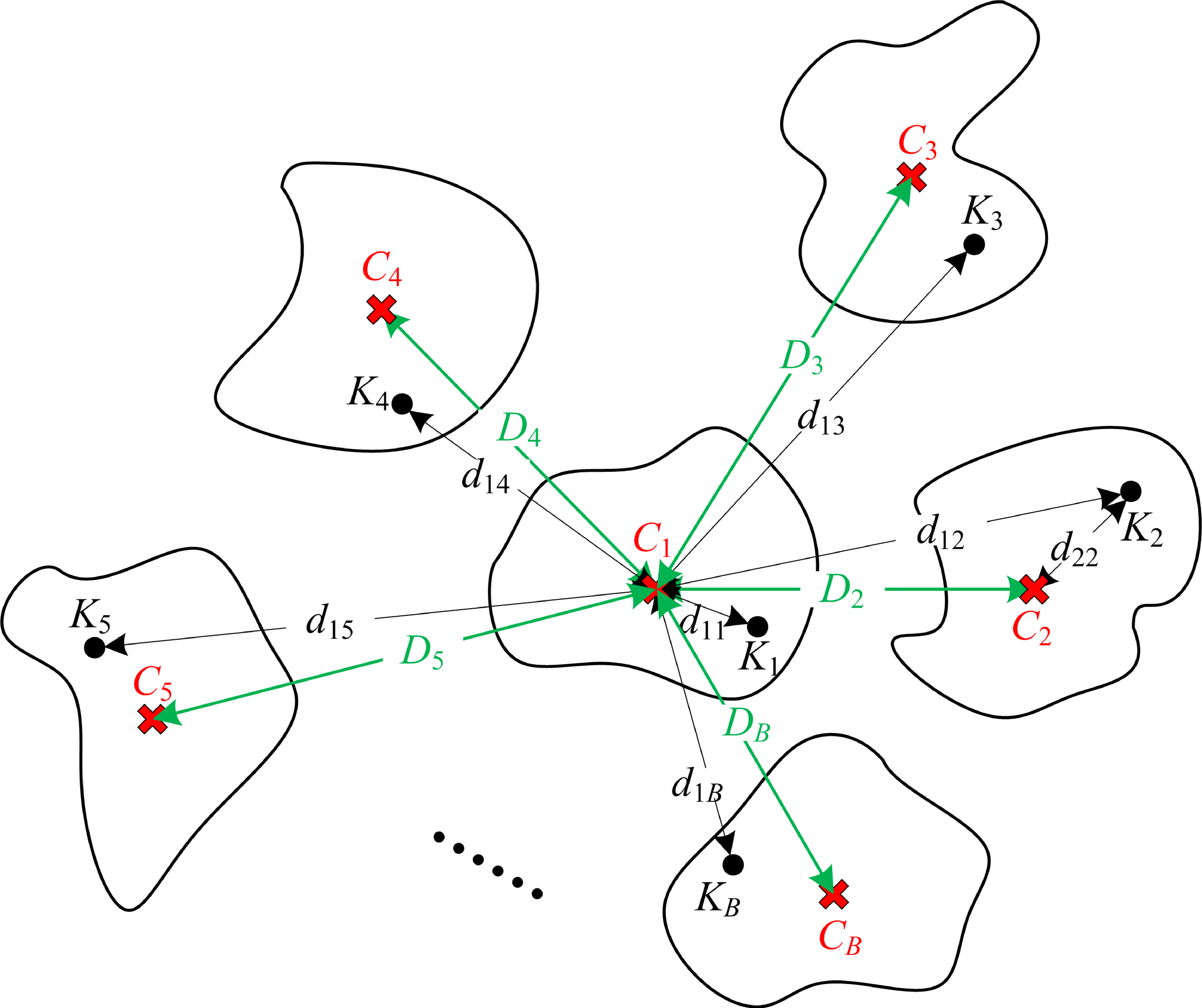} \renewcommand{\figurename}{Fig.}\protect\caption{\label{fig:sys_model}A schematic model of the considered SCN.}

\par\end{centering}

\vspace{-0.8cm}
\end{figure}

\par\end{center}

\textit{The distance} (in km) from the BS of $C_{b}$ to the BS of
$C_{1}$, $b\in\left\{ 1,\ldots,B\right\} $ and the distance from
UE $K_{b}$ to the BS of $C_{m}$, $b,m\in\left\{ 1,\ldots,B\right\} $
are denoted by $D_{b}$ and $d_{bm}$, respectively. In this paper,
we consider a deterministic deployment of BSs, and thus $\left\{ D_{b}\right\} $
is assumed to be known%
. However, the position of UE $K_{b}$ is assumed to be randomly and
uniformly distributed in $R_{b}$. Hence, $d_{bm}$ is a random variable
(RV), the distribution of which cannot be readily expressed in analytical
form due to the arbitrary shape of $R_{b}$.

Based on the definition of $d_{bm}$, \textit{the path loss} (in dB)
from UE $K_{b}$ to the BS of $C_{m}$ is modeled as

\vspace{0.1cm}

\begin{singlespace}
\noindent
\begin{equation}
L_{bm}={A}+{\alpha}{\log_{10}}{d_{bm}},\label{eq:PL_BS2UE}
\end{equation}

\end{singlespace}

\noindent where $A$ and $\alpha$ are the path loss at the reference
distance $d_{bm}=1$ and the path loss exponent, respectively. Note
that $L_{bm}$ is a RV due to the randomness of $d_{bm}$. In practice,
$A$ and $\alpha$ are constants obtainable from field tests~\cite{TR36.828}.

\noindent %

\noindent %

\begin{singlespace}
\noindent %

\end{singlespace}

\textit{The shadow fading} (in dB) from UE $K_{b}$ to the BS of $C_{m}$
is denoted by $S_{bm}$ $\left(b,m\in\left\{ 1,\ldots,B\right\} \right)$,
and it is usually modeled as a zero-mean Gaussian RV because the linear
value of $S_{bm}$ is commonly assumed to follow a lognormal distribution~\cite{TR36.828}.
Hence, in this paper, we model $S_{bm}$ as an independently identical
distributed (i.i.d.) zero-mean Gaussian RV with variance $\sigma_{\textrm{Shad}}^{2}$,
expressed as $S_{bm}\sim\mathcal{N}\left(0,\sigma_{\textrm{Shad}}^{2}\right)$%
\footnote{A more practical assumption would be the correlated shadow fading~\cite{Corr_shadow_fading},
which constructs $S_{bm}$ and $S_{jm}$ $\left(b,j,m\in\left\{ 1,\ldots,B\right\} ,\, b\neq j\right)$
as correlated RVs, and the correlation coefficient should decrease
with the increase of the distance from UE $K_{b}$ to UE $K_{j}$.
Such assumption of the correlated shadow fading will greatly complicate
the analysis since it is difficult to characterize the distribution
of the inter-UE distance. For the sake of tractability, in this paper,
we assume i.i.d. shadow fading for the UE-to-BS links.%
}.

\textit{The multi-path fading} channel vector from UE $K_{b}$ to
the BS of $C_{m}$ is denoted by ${\bf {h}}_{bm}\in\mathbb{C}^{N\times1}$,
where we assume that each UE and each BS are equipped with 1 and $N$
omni-directional antennas, respectively. All channel coefficients
are assumed to experience uncorrelated flat Rayleigh fading, and they
are modeled as i.i.d. zero-mean circularly symmetric complex Gaussian
(ZMCSCG) RVs with unit variance.

\emph{The normalized maximal ratio combining (MRC) reception filter}
denoted by ${{\bf {f}}_{m}}=\frac{{\bf {h}}_{mm}^{\textrm{H}}}{{\left\Vert {\bf {h}}_{mm}\right\Vert }}$
is considered at the BS of $C_{m}$, where ${\bf {h}}_{mm}^{\textrm{H}}$
is the Hermitian transpose of ${\bf {h}}_{mm}$. According to~\cite{MiSA_dynTDD},
the effective channel gain associated with the link from UE $K_{b}$
to the BS of $C_{m}$ $\left(b,m\in\left\{ 1,\ldots,B\right\} ,\, b\neq m\right)$,
defined as $\left|{\bf {f}}_{m}{\bf {h}}_{bm}\right|^{2}$ and denoted
by $H_{bm}$, follows an i.i.d. exponential distribution $\textrm{exp}\left(1\right)$
with the mean equal to 1. %

\textit{The transmit power} of UE $K_{b}$ is denoted by $P_{b}$.
In practice, $P_{b}$ is usually subject to a semi-static power control
(PC) mechanism, e.g., the fractional pathloss compensation (FPC) scheme~\cite{TR36.828}.
Specifically, based on this FPC scheme, $P_{b}$ in dBm is modeled
as%
\footnote{In practice, $P_{b}$ is also constrained by the maximum value of
the UL power and affected by the per-UE signal-to-interference-plus-noise
ratio (SINR) target. The power constraint is a minor issue for UEs
in SCNs since they are generally not power-limited due to the close
proximity of the UEs and the associated SCN BSs~\cite{TR36.828}.
The per-UE SINR target, on the other hand, will greatly complicate
the analysis since it is difficult to model the distribution of the
target SINRs. For the sake of tractability, in this paper, we model
$P_{b}$ as (\ref{eq:UL_P_UE}), which is widely adopted in the literature
{[}3,6-8,11{]}.%
}%

\noindent
\begin{equation}
P_{b}={P_{0}}+\eta\left({L_{bb}+S_{bb}}\right),\label{eq:UL_P_UE}
\end{equation}
where $P_{0}$ is the power basis in dBm on the considered frequency
RB, $\eta\in\left(0,1\right]$ is the FPC factor, $L_{bb}$ has been
defined in (\ref{eq:PL_BS2UE}), and $S_{bb}\sim\mathcal{N}\left(0,\sigma_{\textrm{Shad}}^{2}\right)$.

For clarity, the defined RVs in our system model are summarized in
Table~\ref{tab:RV_def}.

\noindent \begin{center}
\begin{table}
\begin{centering}
{\small{}\protect\caption{\label{tab:RV_def}Definition of RVs.}
}
\par\end{centering}{\small \par}

{\small{}\vspace{-0.1cm}
}{\small \par}

\begin{centering}
{\small{}}%
\begin{tabular}{|l|l|l|}
\hline
{\small{}RV} & {\small{}Description} & {\small{}Distribution}\tabularnewline
\hline
\hline
{\small{}$d_{bm}$} & {\small{}The distance from $K_{b}$ to $C_{m}$} & {\small{}related to $R_{b}$}\tabularnewline
\hline
{\small{}$L_{bm}$} & {\small{}The path loss from $K_{b}$ to $C_{m}$} & {\small{}related to $R_{b}$}\tabularnewline
\hline
{\small{}$S_{bm}$} & {\small{}The shadow fading from $K_{b}$ to $C_{m}$} & {\small{}i.i.d. $\mathcal{N}\left(0,\sigma_{\textrm{Shad}}^{2}\right)$}\tabularnewline
\hline
{\small{}$H_{bm}$} & {\small{}The channel gain from $K_{b}$ to $C_{m}$} & {\small{}i.i.d. $\textrm{exp}\left(1\right)$}\tabularnewline
\hline
{\small{}$P_{b}$} & {\small{}The UL transmission power of $K_{b}$} & {\small{}related to $R_{b}$}\tabularnewline
\hline
\end{tabular}
\par\end{centering}{\small \par}

\vspace{-0.5cm}
\end{table}

\par\end{center}

\section{Analysis of UL Interference\label{sec:Analysis-of-UL-Interf}}

Based on the definition of RVs listed in Table~\ref{tab:RV_def},
\emph{the UL received interference power} (in dBm) from UE $K_{b}$
to the BS of $C_{1}$ can be written as

\noindent
\begin{eqnarray}
I_{b} & \hspace{-0.3cm}\overset{(a)}{=}\hspace{-0.3cm} & P_{b}-L_{b1}-S_{b1}+10\log_{10}H_{b1}\nonumber \\
 & \hspace{-0.3cm}=\hspace{-0.3cm} & P_{0}+\left(\eta L_{bb}-L_{b1}\right)+\left(\eta S_{bb}-S_{b1}\right)+10\log_{10}H_{b1}\nonumber \\
 & \hspace{-0.3cm}\overset{\triangle}{=}\hspace{-0.3cm} & P_{0}+L+S+10\log_{10}H_{b1},\label{eq:rx_interf_I1b_UL}
\end{eqnarray}

\noindent where~(\ref{eq:UL_P_UE}) is plugged into step (a) of~(\ref{eq:rx_interf_I1b_UL}).
Besides, $L$ and $S$ are defined as $L\overset{\triangle}{=}\left(\eta L_{bb}-L_{b1}\right)$
and $S\overset{\triangle}{=}\left(\eta S_{bb}-S_{b1}\right)$, respectively.
In particular, %
since $S_{bb}$ and $S_{b1}$ $\left(b\in\left\{ 2,\ldots,B\right\} \right)$
are i.i.d. zero-mean Gaussian RVs, it is easy to show that $S$ is
also a Gaussian RV with a distribution $\mathcal{N}\left(\mu_{S},\sigma_{S}^{2}\right)$,
where

\vspace{-0.2cm}

\begin{equation}
\begin{cases}
\mu_{S}=0 & \hspace{-0.3cm}\\
\sigma_{S}^{2}=\left(1+\eta^{2}\right)\sigma_{\textrm{Shad}}^{2} & \hspace{-0.3cm}
\end{cases}.\label{eq:comb_shadowing_mean_and_var}
\end{equation}

\vspace{-0.1cm}

From the definition of \emph{$I_{b}$} in~(\ref{eq:rx_interf_I1b_UL}),\emph{
the aggregated interference power} (in mW) from all interfering UEs
to the BS of $C_{1}$ is

\noindent
\begin{equation}
I^{\textrm{mW}}=\sum\limits _{b=2}^{B}{10^{\frac{1}{10}I_{b}}}.\label{eq:rx_interf_UL}
\end{equation}

In the following, we analyze the distribution of $I^{\textrm{mW}}$
in three steps. First, we investigate the distribution of $\left(S+10\log_{10}H_{b1}\right)$
shown in~(\ref{eq:rx_interf_I1b_UL}). Second, we approximate the
distribution of $I_{b}$ as a Gaussian distribution. Third, we show
that the distribution of $I^{\textrm{mW}}$ can be further approximated
as a lognormal distribution.

\subsection{The Distribution of $\left(S+10\log_{10}H_{b1}\right)$ in~(\ref{eq:rx_interf_I1b_UL})\label{sub:approx_LNxEXP}}

According to~\cite{Approx_LNmEXP}, the product of a lognormal RV
and an exponential RV can be well approximated by another lognormal
RV. Therefore, in our case, the sum $\left(S+10\log_{10}H_{b1}\right)$
can be well approximated by a Gaussian RV $G$, because $10^{\frac{1}{10}S}$
is a lognormal RV and $H_{b1}\sim\textrm{exp}\left(1\right)$. %
The mean and variance of $G$ can be respectively computed as~\cite{Approx_LNmEXP}

\vspace{-0.2cm}

\begin{equation}
\begin{cases}
\mu_{G}={\mu_{S}}+\mu_{\textrm{offset}} & \hspace{-0.3cm}\\
\ensuremath{\sigma_{G}^{2}}=\sigma_{S}^{2}+\sigma_{\textrm{offset}}^{2} & \hspace{-0.3cm}
\end{cases}.\label{eq:approx_LNmEXP_mean_and_var_offset}
\end{equation}

\vspace{-0.1cm}

In~\cite{Approx_LNmEXP}, $\mu_{\textrm{offset}}$ and $\sigma_{\textrm{offset}}$
are suggested to be $-2.5$ and $5.57$, respectively. Note that the
approximation is very accurate when $\sigma_{S}^{2}>36$~\cite{Approx_LNmEXP}.
Such requirement is readily satisfied in practical SCNs, e.g., it
is recommended in~\cite{TR36.828} that $\sigma_{\textrm{Shad}}^{2}=100$
and hence $\sigma_{S}^{2}>100>36$ because of (\ref{eq:comb_shadowing_mean_and_var}).

\vspace{-0.1cm}

\subsection{The Distribution of $I_{b}$ in~(\ref{eq:rx_interf_I1b_UL})\label{sub:approx_Ib}}

In the following, we will prove that under a certain condition, $I_{b}$
can be well approximated by a Gaussian RV. First, based on the approximation
discussed in Subsection~\ref{sub:approx_LNxEXP}, we reform~(\ref{eq:rx_interf_I1b_UL})
into

\vspace{-0.3cm}

\begin{equation}
I_{b}\approx P_{0}+L+G.\label{eq:Ib_reform}
\end{equation}

\vspace{-0.1cm}

\noindent %

\noindent Note that the probability density function (PDF) and cumulative
density function (CDF) of $L$ are generally not tractable because
$L$ is a RV with respect to $d_{bb}$ and $d_{b1}$, which jointly
depend on the arbitrary shape %
of $R_{b}$.

Next, we analyze the distribution of $I_{b}$ by investigating the
condition under which the sum of a Gaussian RV and an arbitrary RV,
i.e., $\left(L+G\right)$, can be well approximated by another Gaussian
RV. To that end, we denote by $\mu_{L}$ and $\sigma_{L}^{2}$ the
mean and variance of $L$, respectively, define a zero-mean RV as
$\tilde{L}=L-\mu_{L}$, and further define another zero-mean RV as
$\tilde{G}=G-\mu_{G}$. As a result, (\ref{eq:Ib_reform}) can be
re-written as

\vspace{-0.2cm}

\begin{equation}
I_{b}\approx\ensuremath{\tilde{L}+\tilde{G}+\left(P_{0}+\mu_{L}+\mu_{G}\right)}.\label{eq:Ib_reform_rewritten}
\end{equation}

\subsubsection{The distribution of $\left(\tilde{L}+\tilde{G}\right)$ in~(\ref{eq:Ib_reform_rewritten})}

$\,$

Obviously, if $\left(\tilde{L}+\tilde{G}\right)$ in~(\ref{eq:Ib_reform_rewritten})
can be well approximated by a Gaussian RV, then $I_{b}$ can also
be well approximated by the same Gaussian RV with an offset $\left(P_{0}+\mu_{L}+\mu_{G}\right)$.
In order to prove that $\left(\tilde{L}+\tilde{G}\right)$ indeed
can be well approximated by a Gaussian RV, we introduce the Berry\textendash Esseen
theorem in Theorem~\ref{thm:The-Berry=002013Esseen-theorem:}.
\begin{thm}
\label{thm:The-Berry=002013Esseen-theorem:}{[}Theorem~1, 15{]} The
Berry\textendash Esseen theorem: Let $X_{1},X_{2},\ldots,X_{M}$ be
$M$ independent RVs with $\mathbb{E}\left\{ X_{i}\right\} =0$, $\mathbb{E}\left\{ X_{i}^{2}\right\} =\sigma_{i}^{2}>0$,
and $\mathbb{E}\left\{ \left|X_{i}\right|^{3}\right\} <\infty$, $i\in\left\{ 1,2,\ldots,M\right\} $.
Also, let $X=\frac{\sum_{i=1}^{M}X_{i}}{\sqrt{\sum_{i=1}^{M}\sigma_{i}^{2}}}$.
Denote by $F_{X}\left(x\right)$ and $\Phi\left(x\right)$ the CDF
of $X$ and the CDF of the standard normal distribution. Then, we
have

\noindent
\begin{equation}
\underset{x\in\mathbb{R}}{\sup}\left|F_{X}\left(x\right)-\Phi\left(x\right)\right|\leq C_{0}\psi_{0},\label{eq:the Berry=002013Esseen theorem}
\end{equation}
where $\psi_{0}=\frac{\sum_{i=1}^{M}\mathbb{E}\left\{ \left|X_{i}\right|^{3}\right\} }{\left(\sqrt{\sum_{i=1}^{M}\sigma_{i}^{2}}\right)^{3}}$
and $C_{0}=0.56$.
\end{thm}

In Theorem~\ref{thm:The-Berry=002013Esseen-theorem:}, $C_{0}\psi_{0}$
is the upper bound of the maximum gap between $F_{X}\left(x\right)$
and $\Phi\left(x\right)$. Therefore, Theorem~\ref{thm:The-Berry=002013Esseen-theorem:}
shows \emph{how good} the sum of $M$ independent RVs can be approximated
by a Gaussian RV. As long as $C_{0}\psi_{0}$ is reasonably small,
we can say that the approximation is good. Based on Theorem~\ref{thm:The-Berry=002013Esseen-theorem:},
we propose Lemma~\ref{lem:G2_normal_approx} to address the question:
Under what condition can $\left(\tilde{L}+\tilde{G}\right)$ in~(\ref{eq:Ib_reform_rewritten})
be well approximated by a Gaussian RV?
\begin{lem}
\label{lem:G2_normal_approx}Considering the zero-mean RV $\left(\tilde{L}+\tilde{G}\right)$
given by~(\ref{eq:Ib_reform_rewritten}), we define a normalized
RV as $Y=\frac{\tilde{L}+\tilde{G}}{\sqrt{\sigma_{L}^{2}+\sigma_{G}^{2}}}$.
Denote by $F_{Y}\left(y\right)$ and $\Phi\left(y\right)$ the CDF
of $Y$ and the CDF of the standard normal distribution. Then, we
have

\noindent
\begin{equation}
\underset{y\in\mathbb{R}}{\sup}\left|F_{Y}\left(y\right)-\Phi\left(y\right)\right|\leq\tau,\label{eq:ineq_lemma_3}
\end{equation}

\noindent where $\tau=\frac{C_{0}\mathbb{E}\left\{ \left|\tilde{L}\right|^{3}\right\} }{\left(\sqrt{\sigma_{L}^{2}+\sigma_{G}^{2}}\right)^{3}}$
and $C_{0}=0.56$.\end{lem}
\begin{IEEEproof}
In order to make use of Theorem~\ref{thm:The-Berry=002013Esseen-theorem:}
to prove Lemma~\ref{lem:G2_normal_approx}, we construct $M-1$ zero-mean
Gaussian RVs by breaking the zero-mean Gaussian RV $\tilde{G}$ into
the sum of $M-1$ i.i.d. zero-mean Gaussian RVs $\left\{ G_{i}\right\} ,i\in\left\{ 1,2,\ldots,M-1\right\} $,
i.e., $\tilde{G}=\sum_{i=1}^{M-1}G_{i}$. The variance of each $G_{i}$
is $\sigma_{G_{i}}^{2}=\frac{\sigma_{G}^{2}}{M-1}$. Hence, the constructed
$M-1$ zero-mean Gaussian RVs $\left\{ G_{i}\right\} $ and the RV
$\tilde{L}$ form the set of $M$ i.i.d. RVs required to invoke Theorem~\ref{thm:The-Berry=002013Esseen-theorem:}.
Denote the normalized sum of the $M$ RVs as $Y=\frac{\tilde{L}+\sum_{i=1}^{M-1}G_{i}}{\sqrt{\sigma_{L}^{2}+\sum_{i=1}^{M-1}\sigma_{G_{i}}^{2}}}$.
Then, we can evaluate the upper bound of the maximum gap between $F_{Y}\left(y\right)$
and $\Phi\left(y\right)$ by checking the metric $C_{0}\psi_{0}$
using Theorem~\ref{thm:The-Berry=002013Esseen-theorem:}.

According to Theorem~\ref{thm:The-Berry=002013Esseen-theorem:},
the metric $C_{0}\psi_{0}$ for $Y$ becomes

\vspace{-0.1cm}

\noindent
\begin{eqnarray}
C_{0}\psi_{0} & \hspace{-0.3cm}\overset{(a)}{=}\hspace{-0.3cm} & C_{0}\times\frac{\mathbb{E}\left\{ \left|\tilde{L}\right|^{3}\right\} +\sum_{i=1}^{M-1}\mathbb{E}\left\{ \left|G_{i}\right|^{3}\right\} }{\left(\sqrt{\sigma_{L}^{2}+\sum_{i=1}^{M-1}\sigma_{G_{i}}^{2}}\right)^{3}}\nonumber \\
 & \hspace{-0.3cm}=\hspace{-0.3cm} & C_{0}\hspace{-0.1cm}\times\hspace{-0.1cm}\frac{\mathbb{E}\left\{ \left|\tilde{L}\right|^{3}\right\} +\frac{\left(M-1\right)\left(\sqrt{2}\right)^{3}\Gamma\left(2\right)}{\sqrt{\pi}}\hspace{-0.1cm}\left(\sqrt{\frac{\sigma_{G}^{2}}{M-1}}\right)^{3}}{\left(\sqrt{\sigma_{L}^{2}+\sigma_{G}^{2}}\right)^{3}},\hspace{0.5cm}\label{eq:checking_lemma_3}
\end{eqnarray}

\vspace{-0.2cm}

\noindent where $\mathbb{E}\left\{ \left|G_{i}\right|^{3}\right\} =\frac{\left(\sqrt{2}\right)^{3}\Gamma\left(2\right)}{\sqrt{\pi}}\left(\sqrt{\frac{\sigma_{G}^{2}}{M-1}}\right)^{3}$~\cite{Book_Proakis}
has been plugged into step (a) of~(\ref{eq:checking_lemma_3}). Since
the value of $M$ is arbitrary in Theorem~\ref{thm:The-Berry=002013Esseen-theorem:},
we let $M$ approach infinity%
, and thus the first term of the numerator of $\psi_{0}$ in~(\ref{eq:checking_lemma_3}),
i.e., $\frac{\left(M-1\right)\left(\sqrt{2}\right)^{3}\Gamma\left(2\right)}{\sqrt{\pi}}\hspace{-0.1cm}\left(\sqrt{\frac{\sigma_{G}^{2}}{M-1}}\right)^{3}$,
will diminish to zero and as a result $C_{0}\psi_{0}$ converges to $\tau$ given
by

\vspace{-0.4cm}

\begin{equation}
\tau=\frac{C_{0}\mathbb{E}\left\{ \left|\tilde{L}\right|^{3}\right\} }{\left(\sqrt{\sigma_{L}^{2}+\sigma_{G}^{2}}\right)^{3}}.\label{eq:tao_lemma_3}
\end{equation}

\vspace{-0.1cm}

Our proof is thus completed.
\end{IEEEproof}

\vspace{0.2cm}

An important note on the proof of Lemma~\ref{lem:G2_normal_approx}
is that the decomposition of $\tilde{G}$ into the sum of $M-1$ i.i.d.
zero-mean Gaussian RVs $\left\{ G_{i}\right\} $ allows the diminishing
of $\sum_{i=1}^{M-1}\mathbb{E}\left\{ \left|G_{i}\right|^{3}\right\} $,
which quickly reduces $C_{0}\psi_{0}$ in~(\ref{eq:checking_lemma_3})
to $\tau$, thus making it much easier for the Gaussian approximation
to be valid. %
According to Theorem~\ref{thm:The-Berry=002013Esseen-theorem:},
if $\tau$ in~(\ref{eq:tao_lemma_3}) takes a reasonably small value,
then $\left(\tilde{L}+\tilde{G}\right)$ can be well approximated
by a Gaussian RV.

From Lemma~\ref{lem:G2_normal_approx}, we can see that $\tau$ decreases
with the decrease of {\small{}$\mathbb{E}\left\{ \left|\tilde{L}\right|^{3}\right\} $}
and with the increase of $\sigma_{G}^{2}$ and $\sigma_{L}^{2}$.
Therefore, if the $\frac{\mathbb{E}\left\{ \left|\tilde{L}\right|^{3}\right\} }{\sigma_{L}^{2}}$
associated with UE distribution area $R_{b}$ is low, then $\tau$
tends to be a small value, and hence the approximation holds. Intuitively
speaking, if $\frac{\mathbb{E}\left\{ \left|\tilde{L}\right|^{3}\right\} }{\sigma_{L}^{2}}$
is low, then the interfering UEs will be basically concentrated in
a small area of $R_{b}$. As a result, the geometrical randomness
of the interfering UEs is reduced, and hence the dB-scale shadow fading,
usually modeled as Gaussian distributions, will dominate the dB-scale
UL interference distribution. %
Also, a larger $\sigma_{G}^{2}$ allows a better approximation due
to the more dominance of the Gaussian distribution of the dB-scale
shadow fading. It is important to note that although the PDF and CDF
of $L$ and $\tilde{L}$ are difficult to obtain, if not impossible,
the values of $\sigma_{L}^{2}$, {\small{}$\mathbb{E}\left\{ \left|\tilde{L}\right|^{3}\right\} $}
can be easily computed using numerical integration over the UE distribution
area $R_{b}$. We will briefly discuss the calculation of $\tau$
in the next subsection.

To sum up, according to Lemma~\ref{lem:G2_normal_approx}, if $\tau$
is reasonably small for the considered SCN, e.g., around 0.01 (an
error about 1 percentile), $\left(\tilde{L}+\tilde{G}\right)$ can
be well approximated by a zero-mean Gaussian RV. Then, $I_{b}$ in~(\ref{eq:Ib_reform_rewritten})
can be well approximated by a Gaussian RV $Q_{b}$, the mean and the
variance of which can be computed by

\vspace{-0.2cm}

\begin{equation}
\begin{cases}
\mu_{Q_{b}}=P_{0}+\mu_{L}+\mu_{G} & \hspace{-0.3cm}\\
\ensuremath{\sigma_{Q_{b}}^{2}}=\sigma_{L}^{2}+\sigma_{G}^{2} & \hspace{-0.3cm}
\end{cases}.\label{eq:approx_LxLNxEXP_mean_and_var}
\end{equation}

\subsubsection{The calculation of $\tau$}

$\,$

Considering the definition of $L$, $\tilde{L}$ respectively in~(\ref{eq:rx_interf_I1b_UL})
and~(\ref{eq:Ib_reform_rewritten}), we can directly evaluate $\tau$ using~(\ref{eq:tao_lemma_3})
based on the results of $\mu_{L}$, $\sigma_{L}^{2}$ and {\small{}$\mathbb{E}\left\{ \left|\tilde{L}\right|^{3}\right\} $}
computed by
\small

\noindent \vspace{-0.3cm}

\noindent
\begin{eqnarray}
\mu_{L} & \hspace{-0.3cm}=\hspace{-0.3cm} & \int_{R_{b}}LdZ\nonumber \\
 & \hspace{-0.3cm}=\hspace{-0.3cm} & \int_{R_{b}}\left(\eta L_{bb}-L_{b1}\right)dZ\nonumber \\
 & \hspace{-0.3cm}=\hspace{-0.3cm} & \int_{R_{b}}\left(\left(\eta-1\right)A+{\alpha}{\log_{10}}{\frac{d_{bb}^{\eta}}{d_{b1}}}\right)dZ.\label{eq:mean_Lb_com_special_case}
\end{eqnarray}

\noindent \vspace{-0.3cm}

\noindent
\begin{eqnarray}
\sigma_{L}^{2} & \hspace{-0.3cm}=\hspace{-0.3cm} & \int_{R_{b}}\left(L-\mu_{L}\right)^{2}dZ\nonumber \\
 & \hspace{-0.3cm}=\hspace{-0.3cm} & \int_{R_{b}}\left(\left(\eta-1\right)A+{\alpha}{\log_{10}}{\frac{d_{bb}^{\eta}}{d_{b1}}}-\mu_{L}\right)^{2}dZ.\hspace{0.5cm}\label{eq:var_Lb_com_special_case}
\end{eqnarray}

\noindent \vspace{-0.3cm}

\noindent
\begin{eqnarray}
\mathbb{E}\left\{ \left|\tilde{L}\right|^{3}\right\}  & \hspace{-0.3cm}=\hspace{-0.4cm} & \int_{R_{b}}\hspace{-0.1cm}\left|L-\mu_{L}\right|^{3}dZ\nonumber \\
 & \hspace{-0.3cm}=\hspace{-0.4cm} & \int_{R_{b}}\hspace{-0.1cm}\left|\left(\eta-1\right)A+{\alpha}{\log_{10}}{\frac{d_{bb}^{\eta}}{d_{b1}}}-\mu_{L}\right|^{3}\hspace{-0.2cm}dZ.\hspace{0.5cm}\label{eq:3rd_moment_Lb_com_special_case}
\end{eqnarray}

\noindent \vspace{-0.3cm}

\normalsize

\vspace{-0.1cm}

\subsection{The Distribution of $I^{\textrm{mW}}$ in~(\ref{eq:rx_interf_UL})\label{sub:approx_I_mw}}

According to~\cite{Approx_sumLN}, the sum of multiple independent
lognormal RVs can be well approximated by a lognormal RV. Considering
the expression of $I^{\textrm{mW}}$ in~(\ref{eq:rx_interf_UL}),
if each $I_{b},b\in\left\{ 2,\dots,B\right\} $ can be approximated
by the Gaussian RV $Q_{b}$, we can conclude that $I^{\textrm{mW}}$
can be well approximated by a lognormal RV, denoted by $\hat{I}^{\textrm{mW}}=10^{\frac{1}{{10}}Q}$.
The RV $Q$ is a Gaussian RV and its mean and the variance are denoted
by $\mu_{Q}$ and $\sigma_{Q}^{2}$, respectively. According to~\cite{Approx_sumLN},
$\mu_{Q}$ and $\sigma_{Q}^{2}$ are obtained by solving the following
equation set,

\noindent \vspace{-0.6cm}

\begin{equation}
\left\{ \begin{array}{l}
{{\hat{\Psi}}_{Q}}\left({s_{1}}\right)=\prod\limits _{b=2}^{B}{{{\hat{\Psi}}_{{Q_{b}}}}\left({s_{1}}\right)}\overset{\Delta}{=}C_{1}\\
{{\hat{\Psi}}_{Q}}\left({s_{2}}\right)=\prod\limits _{b=2}^{B}{{{\hat{\Psi}}_{{Q_{b}}}}\left({s_{2}}\right)}\overset{\Delta}{=}C_{2}
\end{array}\right.,\label{eq:eqs_para_for_Gsum}
\end{equation}

\noindent where ${\hat{\Psi}}_{X}\left(s\right)$ is the approximated
moment generating function (MGF) evaluated at $s$ for a lognormal
RV defined as $10^{\frac{1}{{10}}X}$. Such approximated MGF is formulated
as

\vspace{-0.2cm}

\begin{equation}
{\hat{\Psi}}_{X}\left(s\right)=\hspace{-0.1cm}\sum\limits _{m=1}^{M_{0}}{\frac{{w_{m}}}{{\sqrt{\pi}}}\exp\left(\hspace{-0.1cm}{-s\exp\hspace{-0.1cm}\left(\hspace{-0.1cm}{\frac{{\sqrt{2\sigma_{X}^{2}}{a_{m}}+{\mu_{X}}}}{\zeta}}\right)}\hspace{-0.1cm}\right)},
\end{equation}

\noindent where $\zeta=\frac{10}{\ln10}$, $M_{0}$ is the order of
the Gauss-Hermite numerical integration, the weights $w_{m}$ and
abscissas $a_{m}$ for $M_{0}$ up to 20 are tabulated in Table 25.10
in~\cite{GH_num_integration}. Usually, $M_{0}$ is larger than 8
to achieve a good approximation~\cite{Approx_sumLN}.

In~(\ref{eq:eqs_para_for_Gsum}), $s_{1}$ and $s_{2}$ are two design
parameters for generating two equations that can determine the appropriate
values of $\mu_{Q}$ and $\sigma_{Q}^{2}$. Generally speaking, when
$s_{1}$ and $s_{2}$ take smaller (larger) values, the mismatch in
the head (tail) portion between the actual CDF and the approximated
one can be reduced~\cite{Approx_sumLN}. The basic idea of~(\ref{eq:eqs_para_for_Gsum})
is to let the two concerned RVs, i.e., $I^{\textrm{mW}}$ and $\hat{I}^{\textrm{mW}}$,
statistically match with each other in the sense of having two equal
points on their MGFs. The solution of~(\ref{eq:eqs_para_for_Gsum})
can be readily found by standard mathematical software programs such
as MATLAB. %

Based on this lognormal approximation, the approximated PDF and CDF
of $I^{\textrm{mW}}$ can be respectively expressed as~\cite{Book_Proakis}

\noindent \vspace{-0.3cm}
\begin{eqnarray}
{f_{I^{\textrm{mW}}}}\left(v\right) & \hspace{-0.3cm}\approx\hspace{-0.3cm} & \frac{\zeta}{{v\sqrt{2\pi\ensuremath{\sigma_{Q}^{2}}}}}\exp\left\{ {-\frac{{{\left({\zeta\ln v-{\mu_{Q}}}\right)}^{2}}}{{2\ensuremath{\sigma_{Q}^{2}}}}}\right\} ,\hspace{0.5cm}\label{eq:PDF_LN_comb_shad}
\end{eqnarray}

\noindent \vspace{-0.4cm}
and

\noindent \vspace{-0.5cm}
\begin{eqnarray}
{F_{I^{\textrm{mW}}}}\left(v\right) & \hspace{-0.3cm}\approx\hspace{-0.3cm} & \frac{1}{2}+\frac{1}{2}\textrm{erf}\left({\frac{{\zeta\ln v-{\mu_{Q}}}}{{\sqrt{2\sigma_{Q}^{2}}}}}\right),\label{eq:CDF_LN_comb_shad}
\end{eqnarray}
where $\textrm{erf}\left(\cdot\right)$ is the error function.

\vspace{-0.3cm}

\subsection{Summary of the Proposed Analysis of UL Interference\label{sub:Summary_approx_analysis}}

To sum up, we highlight the steps in our proposed analysis of UL interference
in the following. First, we use (\ref{eq:mean_Lb_com_special_case}),
(\ref{eq:var_Lb_com_special_case}) and (\ref{eq:3rd_moment_Lb_com_special_case})
to calculate $\mu_{L}$, $\sigma_{L}^{2}$ and {\small{}$\mathbb{E}\left\{ \left|\tilde{L}\right|^{3}\right\} $}
for each $R_{b},b\in\left\{ 2,\ldots,B\right\} $. Then, based on
the calculated results, we check whether $\tau$ given by Lemma~\ref{lem:G2_normal_approx}
is reasonably small. If it is the case, we can approximate $I_{b}$
as a Gaussian RV $Q_{b}$ according to the discussion in Subsection~\ref{sub:approx_Ib}.
Finally, we approximate the RV $I^{\textrm{mW}}$ in~(\ref{eq:rx_interf_UL})
as a lognormal RV $\hat{I}^{\textrm{mW}}=10^{\frac{1}{{10}}Q}$, with
the distribution parameters $\mu_{Q}$ and $\sigma_{Q}^{2}$ obtained
from solving the equation set (\ref{eq:eqs_para_for_Gsum}). The chain
of approximation to obtain $\hat{I}^{\textrm{mW}}$ is summarized
in Table~\ref{tab:approx_chain} shown on the top of next page.

\noindent \begin{center}
\begin{table*}
\begin{centering}
{\small{}\protect\caption{\label{tab:approx_chain}The chain of approximation to obtain $\hat{I}^{\textrm{mW}}$.}
}
\par\end{centering}{\small \par}

{\small{}\vspace{-0.1cm}
}{\small \par}

\begin{centering}
{\small{}}%
\begin{tabular}{|l|l|l|l|}
\hline
{\small{}Original RV} & {\small{}Approximated RV} & {\small{}Approximated distribution} & {\small{}Distribution parameters}\tabularnewline
\hline
\hline
{\small{}$S+10\log_{10}H_{b1}$ in~(\ref{eq:rx_interf_I1b_UL})} & {\small{}$G$} & {\small{}Gaussian} & {\small{}See (\ref{eq:approx_LNmEXP_mean_and_var_offset})}\tabularnewline
\hline
{\small{}$I_{b}\approx P_{0}+L+G$ in~(\ref{eq:Ib_reform})} & {\small{}$Q_{b}$} & {\small{}Gaussian} & {\small{}See (\ref{eq:approx_LxLNxEXP_mean_and_var})}\tabularnewline
\hline
{\small{}$I^{\textrm{mW}}=\sum_{b=2}^{B}{10^{\frac{1}{10}I_{b}}}$
in~(\ref{eq:rx_interf_UL})} & {\small{}$\hat{I}^{\textrm{mW}}$} & {\small{}Lognormal} & {\small{}The solution of (\ref{eq:eqs_para_for_Gsum})}\tabularnewline
\hline
\end{tabular}
\par\end{centering}{\small \par}

\vspace{0.1cm}

\rule[0.5ex]{2\columnwidth}{0.5pt}

\vspace{-0.5cm}
\end{table*}

\par\end{center}

\section{Simulation and Discussion\label{sec:Simulaiton-and-Discussion}}

In order to validate the results from the proposed micro-scopic analysis
of UL interference, we conduct simulations considering two scenarios,
with a single interfering cell and with multiple interfering cells,
respectively. According to the 3GPP standards~\cite{TR36.828}, the
system parameters in our simulations are set to: $A=103.8$, $\alpha=20.9$,
$P_{0}=-76$\ dBm, $\eta=0.8$, and $\sigma_{S}^{2}=100$. Besides,
the minimum BS-to-UE distance is assumed to be 0.005\ km~\cite{TR36.828}.
Regarding the equation set (\ref{eq:eqs_para_for_Gsum}) to determine
$\mu_{Q}$ and $\sigma_{Q}^{2}$, we choose the parameters as $s_{1}=1$,
$s_{2}=0.1$ and $M_{0}=12$ as recommended in~\cite{Approx_sumLN}.

\subsection{The Scenario with a Single Interfering Cell\label{sub:The-Scenario-single-cell}}

In this scenario, we consider $B=2$, as shown in Fig.~\ref{fig:R2_bread}.
The x-markers indicate BS locations, where the BS location of $C_{1}$
has been explicitly pointed out. The dash-dot line indicate a reference
disk to illustrate the reference size of small cell $C_{2}$. The
radius of such reference circle is denoted by $r$, and the distance
between the BS of $C_{1}$ and the BS of $C_{2}$, i.e., $D_{2}$,
is assumed to be $1.5r$. In our simulations, the values of $r$ (in
km) are set to 0.01, 0.02 and 0.04, respectively. In this scenario,
the interfering UE $K_{2}$ is randomly and uniformly distributed
in an irregularly shaped UE distribution area $R_{2}$, as illustrated
by the area outlined using solid line in Fig.~\ref{fig:R2_bread}.
The shape of $R_{2}$ is the intersection of a square, a circle and
an ellipse, which has a complicated generation function. Examples
of the possible positions of $K_{2}$ within $R_{2}$ are shown as
dots.

\vspace{-0.5cm}

\noindent \begin{center}
\begin{figure}[H]
\noindent \begin{centering}
\vspace{-0.3cm}
\includegraphics[width=6.5cm]{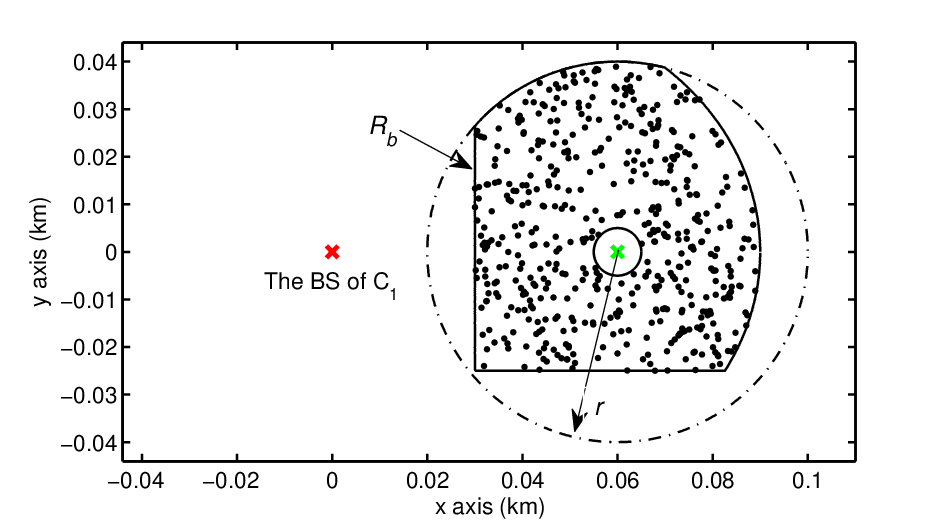}\renewcommand{\figurename}{Fig.}\protect\caption{\label{fig:R2_bread}Illustration of $R_{2}$ (irregular shape, $r=0.04$).}

\par\end{centering}

\vspace{-0.3cm}
\end{figure}

\par\end{center}

\vspace{-0.8cm}

\noindent
\begin{table}[H]
\begin{centering}
{\small{}\protect\caption{\label{tab:results_bread}Results of the proposed analysis ($B=2$).}
}
\par\end{centering}{\small \par}

{\small{}\vspace{-0.3cm}
}{\small \par}

\begin{centering}
{\small{}}%
\begin{tabular}{|l|l|l|l|}
\hline
{\small{}The value of $r$} & {\small{}$\tau$} & {\small{}$\mu_{Q_{2}}$} & {\small{}$\sigma_{Q_{2}}^{2}$}\tabularnewline
\hline
\hline
{\small{}$r=0.01$} & {\small{}0.0082} & {\small{}-97.10} & {\small{}205.25}\tabularnewline
\hline
{\small{}$r=0.02$} & {\small{}0.0122} & {\small{}-99.69} & {\small{}207.73}\tabularnewline
\hline
{\small{}$r=0.04$} & {\small{}0.0184} & {\small{}-101.54} & {\small{}211.44}\tabularnewline
\hline
\end{tabular}
\par\end{centering}{\small \par}

\vspace{-0.2cm}
\end{table}

\vspace{-0.2cm}

Despite of the complicated shape of $R_{2}$, our proposed lognormal
approximation of the UL interference still works. Specifically, from
(\ref{eq:mean_Lb_com_special_case}), (\ref{eq:var_Lb_com_special_case}),
(\ref{eq:3rd_moment_Lb_com_special_case}) and Lemma~\ref{lem:G2_normal_approx},
we can calculate $\mu_{L}$, $\sigma_{L}^{2}$, $\mathbb{E}\left\{ \left|\tilde{L}\right|^{3}\right\} $
and $\tau$, respectively. The results of $\tau$ are tabulated for various values
of $r$ in Table~\ref{tab:results_bread}. From Table~\ref{tab:results_bread},
we can observe that, when $r=0.01$ or $0.02$, the values of $\tau$
are around 0.01. Consequently, it indicates a good approximation of
$I_{2}$ as a Gaussian RV. Note that $r=0.01$ and $r=0.02$ correspond
to the typical network configurations for future dense/ultra-dense
SCNs~\cite{Tutor_smallcell}, which shows that the derived condition
in Lemma~\ref{lem:G2_normal_approx} can be readily used to study
future dense/ultra-dense SCNs.

To confirm the accuracy of the proposed approximation summarized in
Table~\ref{tab:approx_chain}, we plot the approximated analytical
results and the simulation results of $I_{2}$ for the considered
$R_{2}$ in Fig.~\ref{fig:UL_interf_approx_irreg}. Also, the numerical
results of $\mu_{Q_{2}}$ and $\sigma_{Q_{2}}^{2}$ are shown in
Table~\ref{tab:results_bread}. As can be seen from Fig.~\ref{fig:UL_interf_approx_irreg},
the proposed Gaussian approximation of $I_{2}$ is indeed very tight
for the considered $R_{2}$ with such an irregular shape. Note that the
approximation appears to be very good even for $r=0.04$ (despite
a relatively large value of $\tau$ at around 0.02), which implies
that the derived condition in Lemma~\ref{lem:G2_normal_approx} might
be too strict and the proposed approximation might be extended to
more general conditions. Hence, it is our future work to find a way
to relax the sufficient condition proposed in Lemma~\ref{lem:G2_normal_approx}
to allow for a wider application of the proposed analysis.

\vspace{-0.4cm}

\noindent \begin{center}
\begin{figure}
\noindent \begin{centering}
\vspace{-0.3cm}
\includegraphics[width=6.5cm]{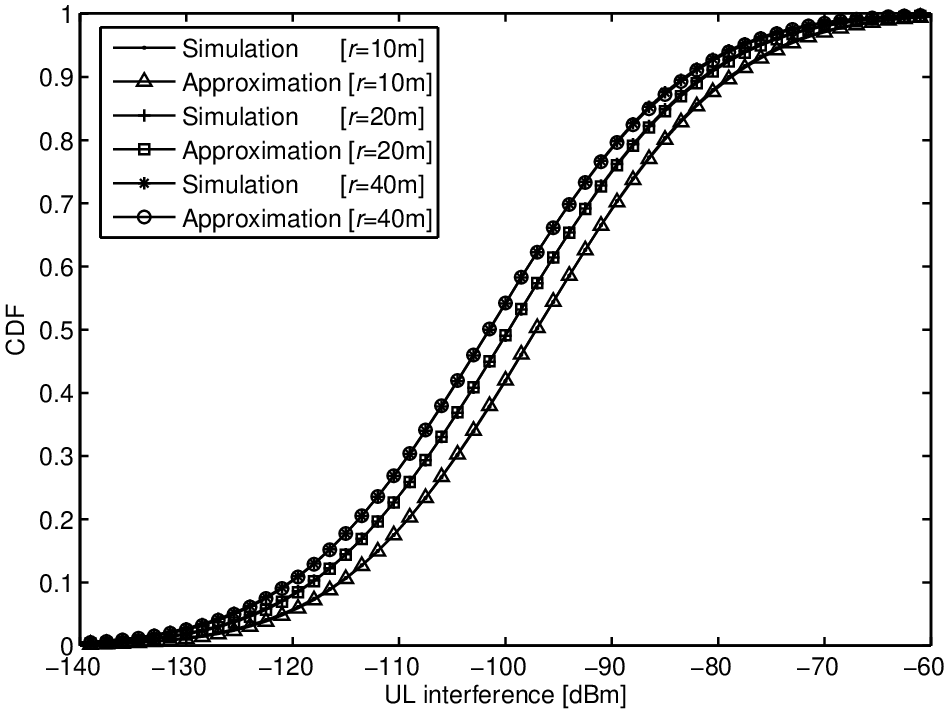}\renewcommand{\figurename}{Fig.}\protect\caption{\label{fig:UL_interf_approx_irreg}The simulation and approximation
of $I_{2}$ (irregular shape of $R_{2}$).}

\par\end{centering}

\vspace{-0.5cm}
\end{figure}

\par\end{center}

\subsection{The Scenario with Multiple Interfering Cells\label{sub:The-Scenario-multiple-cells}}

The lognormal approximation of the UL interference for a network with
\emph{multiple interfering cells placed on a hexagonal lattice} has
been validated in~\cite{UL_interf_LN_conjecture_WCL}. As a significant
leap from the hexagonal network, in this subsection, we apply the
proposed framework on a more complex network with \emph{practical
deployment of multiple cells} and provide the approximation of the
UL interference distribution.

Here, we consider a 3GPP-compliant scenario~\cite{TR36.828}, as
shown in Fig.~\ref{fig:UE_distribution_multicells_hotspot}. In Fig.~\ref{fig:UE_distribution_multicells_hotspot},
$B$ is set to 84 and all small cell BSs are represented by x-markers.
Particularly, the BS of $C_{1}$ has been explicitly pointed out.
The reference UE distribution area for each small cell is a disk with
a radius of $r$~\cite{TR36.828}. In our simulations, the values
of $r$ (in km) are set to 0.01, 0.02 and 0.04, respectively. The
reference disk-shaped areas can be easily seen in Fig.~\ref{fig:UE_distribution_multicells_hotspot}
from any isolated small cell. However, due to the irregular positions
of the cells, the actual UE distribution areas of the considered cells
are of irregular shapes due to cell overlapping. The irregularly shaped
UE distribution areas are outlined in Fig.~\ref{fig:UE_distribution_multicells_hotspot}
by solid lines. Interfering UEs are randomly and uniformly distributed
in those areas. An important note is that the considered network scenario
is different from that adopted in~{[}2,3{]}, where UE distribution
areas are defined as Voronoi cells generated by the Poisson distributed
BSs and those Voronoi cells cover the whole network area. In practice,
small cells are mainly used for capacity boosting in specific populated
areas, rather than provision of an umbrella coverage for all UEs.
Therefore, the 3GPP standards recommend the hotspot SCN scenario depicted
in Fig.~\ref{fig:UE_distribution_multicells_hotspot} for UE distribution
in the performance evaluation of practical SCNs. Nevertheless, the
proposed micro-scopic analysis of UL interference can still be applied
on a particular Voronoi tessellation, since the derived condition
in Lemma~\ref{lem:G2_normal_approx} does not rely on particular
shapes/sizes of UE distribution areas.

\vspace{-0.4cm}

\noindent \begin{center}
\begin{figure}
\noindent \begin{centering}
\vspace{-0.3cm}
\includegraphics[width=6.5cm]{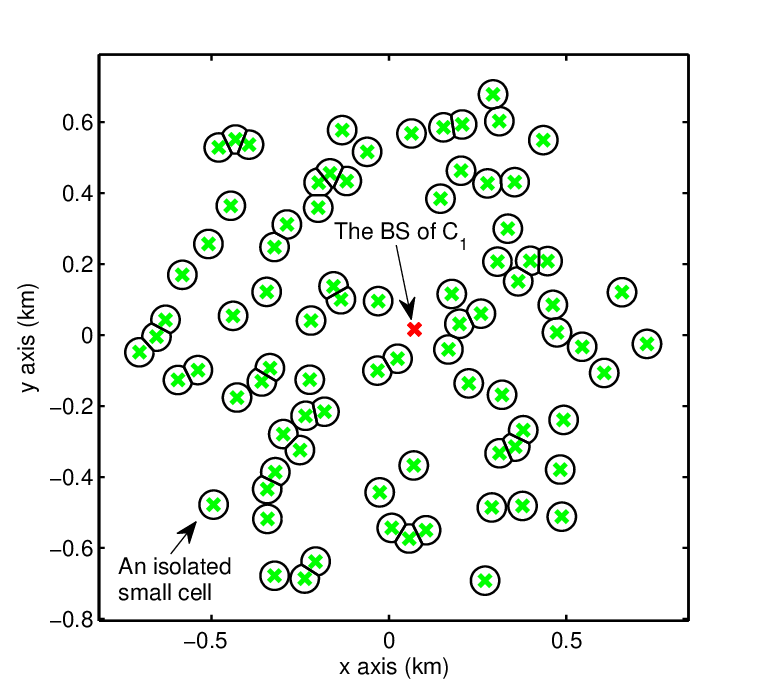}\renewcommand{\figurename}{Fig.}\protect\caption{\label{fig:UE_distribution_multicells_hotspot}Illustration of a practical
deployment of multiple cells ($r=0.04$).}

\par\end{centering}

\vspace{-0.5cm}
\end{figure}

\par\end{center}

\vspace{-0.4cm}

To the best knowledge of the authors, the micro-scopic analysis of
UL interference for such complex network as shown in Fig.~\ref{fig:UE_distribution_multicells_hotspot}
has never been attempted before in the literature. Now, we investigate
the considered network with the proposed approach of performance analysis.
First, we use (\ref{eq:mean_Lb_com_special_case}), (\ref{eq:var_Lb_com_special_case})
and (\ref{eq:3rd_moment_Lb_com_special_case}) to calculate $\mu_{L}$,
$\sigma_{L}^{2}$ and {\small{}$\mathbb{E}\left\{ \left|\tilde{L}\right|^{3}\right\} $}
for \emph{each} $R_{b},b\in\left\{ 2,\ldots,84\right\} $ displayed
in Fig.~\ref{fig:UE_distribution_multicells_hotspot}. Then, based
on the calculated results, we check the value of $\tau$ for \emph{each}
$R_{b}$ using Lemma~\ref{lem:G2_normal_approx}. The maximum values
of the 83 $R_{b}$-specific $\tau$'s for various $r$ values are
presented in Table~\ref{tab:results_multicells_hotspot}. It is shown
in the table that when $r=0.01$ and $r=0.02$ the maximum $\tau$
is below or around 0.01, and thus %
each $I_{b}$ can be approximated by a Gaussian RV $Q_{b}$ according
to the discussion in Subsection~\ref{sub:approx_Ib}. Again, note
that $r=0.01$ and $r=0.02$ correspond to the typical network configurations
for future dense/ultra-dense SCNs~\cite{Tutor_smallcell}, which
indicates the usefulness of the proposed approach of the micro-scopic
analysis in future 5G SCNs. Finally, we approximate the RV $I^{\textrm{mW}}$
in~(\ref{eq:rx_interf_UL}) as a lognormal RV $\hat{I}^{\textrm{mW}}=10^{\frac{1}{{10}}Q}$,
with the distribution parameters $\mu_{Q}$ and $\sigma_{Q}^{2}$
obtained from solving the equation set (\ref{eq:eqs_para_for_Gsum}).
The numerical results of $\mu_{Q}$ and $\sigma_{Q}^{2}$ are provided
in Table~\ref{tab:results_multicells_hotspot}.

\vspace{-0.2cm}

\noindent
\begin{table}[H]
\begin{centering}
{\small{}\protect\caption{\label{tab:results_multicells_hotspot}Results of the proposed analysis
($B=84$).}
}
\par\end{centering}{\small \par}

{\small{}\vspace{-0.4cm}
}{\small \par}

\begin{centering}
{\small{}}%
\begin{tabular}{|l|l|l|l|}
\hline
{\small{}The value of $r$} & {\small{}Maximum $\tau$} & {\small{}$\mu_{Q}$} & {\small{}$\sigma_{Q}^{2}$}\tabularnewline
\hline
\hline
{\small{}$r=0.01$} & {\small{}0.0066} & {\small{}-75.09} & {\small{}17.77}\tabularnewline
\hline
{\small{}$r=0.02$} & {\small{}0.0125} & {\small{}-77.21} & {\small{}18.30}\tabularnewline
\hline
{\small{}$r=0.04$} & {\small{}0.0176} & {\small{}-78.76} & {\small{}18.55}\tabularnewline
\hline
\end{tabular}
\par\end{centering}{\small \par}

\vspace{-0.2cm}
\end{table}

To validate the accuracy of our results on the UL interference, we
plot the approximated analytical results and the simulation results
of $I^{\textrm{mW}}$ in dBm for the considered network in Fig.~\ref{fig:UL_interf_approx_multicells_hotspot}.
As can be seen from Fig.~\ref{fig:UL_interf_approx_multicells_hotspot},
the resulting lognormal approximation of $I^{\textrm{mW}}$ is reasonably
good for practical use. Note that the approximation shown in Fig.~\ref{fig:UL_interf_approx_multicells_hotspot}
is not as tight as that exhibited in Fig.~\ref{fig:UL_interf_approx_irreg}.
The noticeable approximation errors in Fig.~\ref{fig:UL_interf_approx_multicells_hotspot}
are mostly caused by the inaccuracy of approximating the sum of multiple
lognormal RVs as a single lognormal RV in~\cite{Approx_sumLN}. Note
that the parameters of $s_{1}=1$ and $s_{2}=0.1$ recommended by~\cite{Approx_sumLN}
allow for an overall match between the actual CDF and the approximated
lognormal CDF, though the approximation in the head portion and the
tail portion of the CDF are compromised to some extent. Hence, a straightforward
way to improve the approximation of UL interference shown in Fig.~\ref{fig:UL_interf_approx_multicells_hotspot}
is to approximate the CDF of $I^{\textrm{mW}}$ as a piece-wise lognormal
CDFs, because smaller (larger) $s_{1}$ and $s_{2}$ are helpful to
increase the quality of approximation in the head (tail) portion of
the considered CDF~\cite{Approx_sumLN}. Regarding the individual
approximation of $I_{b}$ as a Gaussian RV $Q_{b}$, we find that
it is tight for every considered $R_{b}$. Due to space limitation,
we omit the investigation on the Gaussian approximation for each $I_{b}$,
which is quite similar to the discussion in Subsection~\ref{sub:The-Scenario-single-cell}.

\vspace{-0.2cm}

\noindent \begin{center}
\begin{figure}[H]
\noindent \begin{centering}
\includegraphics[width=6.5cm]{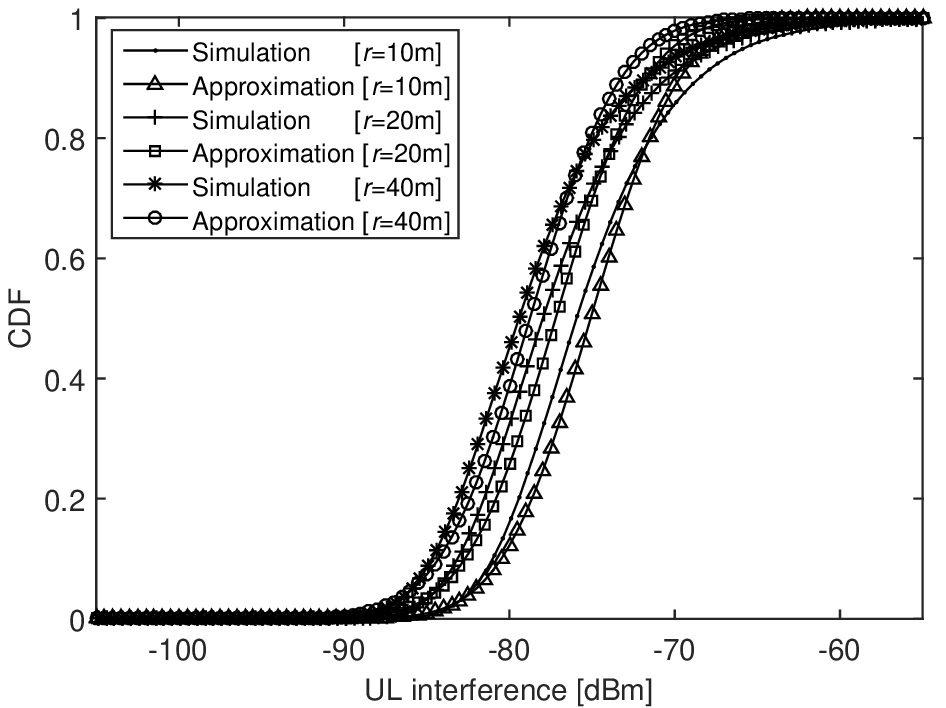}\renewcommand{\figurename}{Fig.}\protect\caption{\label{fig:UL_interf_approx_multicells_hotspot}The simulation and
approximation of $I^{\textrm{mW}}$ in dBm.}

\par\end{centering}

\vspace{-0.7cm}
\end{figure}

\par\end{center}

A final note on the proposed network performance analysis is that
our approach is very powerful to obtain the analytical results in
an efficient manner. The computational complexity is mainly attributable
to the numerical integration to obtain the values of $\mu_{L}$, $\sigma_{L}^{2}$
and {\small{}$\mathbb{E}\left\{ \left|\tilde{L}\right|^{3}\right\} $}
for the 83 interfering cells. In contrast, the simulation involves
a tremendously high complexity. Specifically, in our simulations,
around \emph{one billion} of realizations of $I_{b}$ have been conducted
for the 83 interfering cells in order to go through the randomness
of all the RVs listed in Table~\ref{tab:RV_def}. This shows that
the proposed micro-scopic analysis of network performance is elegant
and computationally efficient, which makes it ideal to study future
5G systems with general and dense small cell deployments.


\section{Conclusion\label{sec:Conclusion}}

The conjecture of approximating the UL inter-cell interference in
FDMA SCNs as a lognormal distribution is vital because it allows tractable
network performance analysis with closed-form expressions. Compared
with the previous works based on empirical studies, our work, for
the first time, analytically proved that the conjecture is conditionally
correct and the derived condition does not rely on particular shapes/sizes
of UE distribution areas. Based on our work, we proposed a new approach
to directly and analytically investigate a complex network with \emph{practical
deployment of multiple cells} based on the approximation of the UL
interference distribution.
The proposed approach has the following merits.
\begin{enumerate}
\item It can deal with any shape of UE distribution area, which tolerates
more practical assumptions, e.g., irregular hot-spots, overlapped
small cells, etc.
\item It measures the quality of approximation using a closed-form expression
and the simulation results validate the tightness of the approximation.
\item It can cope with a
large number of small cells by a low computational
complexity of analysis.
\end{enumerate}

As future work, we will find a way to relax the derived condition
of valid approximation to allow for a wider application of the proposed
analysis.
Besides, we will consider non-uniform UE distribution and other types of multi-path fading, e.g., Rician fading or Nakagami fading, in our analysis.
Also, we will further investigate
alternative distributions of the sum of multiple lognormal RVs to make our analysis more accurate.

\noindent %

\end{document}